# Femtosecond two-photon-excited backward lasing of atomic hydrogen in flame

Pengji Ding,[1,][*] Maria Ruchkina,[1] Yi Liu,[2] Marcus Aldén,[1] and Joakim Bood[1]

[1]*Department of Physics, Lund University, SE-22100 Lund, Sweden*
[2]*Shanghai Key Lab of Modern Optical System, University of Shanghai for Science and Technology, 516, Jungong Road, 200093 Shanghai, China*
[*]*corresponding author:* pengji.ding@forbrf.lth.se

## Abstract

We report on an observation of bi-directional 656 nm lasing action of atomic hydrogen in premixed $CH_4$/air flame induced by resonant femtosecond 205 nm two-photon excitation. In particular, the backward-propagating lasing pulse is systematically characterized in the spectral, spatial and temporal domains for the sake of single-ended diagnostic. Its picosecond-scale duration enables spatially resolved concentration measurements of hydrogen atoms in millimeter range, which is successfully demonstrated using two narrow welding flames.

## Introduction

Lasing (or stimulated emission) via resonant optical excitation of species present in flames is a widespread phenomenon. ASE-type (amplified spontaneous emission) lasing was observed by focusing 5-ns 226-nm laser pulses into $H_2/O_2$ flame at sub-atmospheric pressure and room-temperature flows of $O_2$ and $N_2O$ [1]. Through resonant two-photon excitation, 845 nm lasing of atomic oxygen is generated in both the forward and backward directions. As follows, the lasing effect was also observed in other atoms and molecules such as H [2], C [3], N [4], CO [5] and $NH_3$ [6]. The backward lasing, propagating in the direction opposite to the pump laser beam, is of particularly interest for single-ended combustion diagnostics where only one optical access is available. It should be noted that these earlier works in the field of laser-based combustion diagnostics initiated intensive studies of backward lasing in ambient air during the last decade [7–14], ultimately aiming at remote atmospheric sensing.

The coherent nature of lasing provides some obvious advantages over laser-induced fluorescence (LIF) detection in signal strength and directionality, suggesting that lasing could be a promising diagnostic technique. In addition, lasing techniques are capable of measuring minor species in combustion process that other techniques like, Raman scattering or coherent anti-stokes Raman scattering (CARS), are unable to detect. However, lasing techniques based on nanosecond pump laser pulses also possess some disadvantages. Firstly, generation of the lasing pulse along the pump laser beam as well as its long duration result in a very poor spatial resolution in the direction of the pump beam, allowing only measurements of vertical profiles [1–3,6]. Secondly, photochemical production of lasing species distorts the measured signals and therefore makes the measurements unreliable [2]. Thirdly, the lasing signal strength is not proportional to the number of excited atoms along its path until the saturation effect causes the signal to convert from exponential to linear growth. In other words, it requires sufficiently high laser power.

In this letter, we demonstrate that the aforementioned difficulties of the lasing diagnostic technique can be overcome by using femtosecond laser pulses. With the use of a tunable deep-UV 125 femtosecond (fs) pump laser, a picosecond-time-scale 656 nm lasing pulse of atomic hydrogen was generated in a premixed flame in both the forward and backward directions. As shown in Fig. 1(a), the two-photon excitation transition occurs from the $^1$S ground state to the $^3$D excited state using a 205-nm pump laser. It is followed by relaxation from the $^3$D state to the $^2$P state (the Balmer-α line), releasing lasing emission at 656 nm wavelength (see the spectrum in Fig. 1(b)). An increased spectral bandwidth of the pump laser pulse, compared to nanosecond laser pulses, results in a lasing signal with a much broader detuning range of the pump laser wavelength. The characteristics of the backward 656 nm lasing pulse, including emission spectrum, spatial and temporal profiles, are analyzed. With the backward 656 nm lasing pulses generated from two welding flames, approximately 7 mm spatial resolution for atomic hydrogen detection was achieved, a significant progress towards single-ended diagnostics.

**Experimental setup**

The experimental setup is schematically illustrated in Fig. 1(c). A Chirped Pulsed Amplification (CPA) laser system (Coherent, Hidra-50) is used to deliver 125 fs, 800 nm laser pulse with maximum pulse energy of 30 mJ at 10 Hz repetition rate. This laser beam pumps an Optical Parametric Amplifier (OPA) followed by a frequency mixing apparatus (NirUVis unit), which can provide 205 nm laser pulses with a maximum pulse energy of approximately 50 µJ. The beam diameter is about 5 mm. The 205 nm pump laser propagates through a bulk $CaF_2$ equilateral dispersive prism with an incident angle of 31.6º, in order to spatially separate the backward-propagating 656 nm radiation from the 205 nm pump laser beam. In addition, this configuration spectrally purifies the pump laser by dispersing residual frequencies. After the prism, the pulse energy of the pump beam has been reduced to approximately 20 µJ mainly due to reflection losses. Then, the pump laser beam is focused by a spherical lens ($f$ =300 mm) into a $CH_4$/Air flame and creates a two-photon excitation volume of ~100 µm diameter and 2.0 mm length. A modified porous-plug burner (McKenna) is used (see Fig. 1(d)), which consists of a central tube (2.0 mm inner diameter) and the diameter of the porous plug is 60 mm. Premixed $CH_4$/Air gas mixtures enter the central tube and the porous plug separately through mass flow controllers. The central jet flame is stabilized on the burner by the surrounding flat flame. The equivalence ratio ϕ was 0.9 for the flat flame and 1.3 for the jet flame.

In the forward direction, a bandpass 656 nm filter (Semrock, ~15 nm bandwidth) was used to transmit the induced lasing at 656 nm which was then detected by a calibrated photodiode. In the backward direction, a He-Ne laser operating at 632.8 nm wavelength was employed to roughly determine the separation angle between the backward 656 nm lasing beam and the 205 nm pump beam outside the dispersive prism. For detection, an intensified CCD camera (Princeton Instrument, PI-MAX 2) was used to capture the spatial profile of backward 656 nm lasing beam. A streak camera (Optronis OPTOSCOPE) with capability of 2 ps resolution was used to measure the temporal profile of the lasing pulse. It operated with a streak rate of 10 ps/mm, gain voltage of 880 V, and its entrance slit width was set to 0.18 mm.

**Results and Discussion**

The emission spectrum of the lasing pulse is presented in Fig. 1(b), and it shows a single spectral line centered at a wavelength of 656.064 nm. By collecting the signal with an $f$=100 mm lens and putting white papers in front of the detectors, a red spot can be observed by the naked eye in both the forward and backward directions. Unlike the situation with nanosecond/picosecond pumping, where the lasing strength is roughly equal in the two directions, the forward lasing is much brighter than the backward one in our experiments. This asymmetry in forward versus backward lasing was also observed in lasing of nitrogen molecules and ions in femtosecond laser filamentation, and was found to originate from the short lifetime of the optical gain due to ultrafast excitation and the traveling-wave excitation nature of a pencil-shaped excitation volume [10,11,13,14].

We then studied the pump laser wavelength dependence of the 656 nm lasing signal. The result is presented in Fig. 2. The lasing signal was measured with a calibrated photodiode, and each point is an average for 500 laser shots. For each point, the central wavelength of the pump laser pulse was also measured before the flame. In order to maintain it while detuning the central wavelength of pump laser pulse, the stability of the pump pulse energy was simultaneously monitored by recording the energy of laser light reflected off the incident surface of the dispersive prism, indicated by the open square curve in Fig. 2. It can be seen that the dependence curve ranges from 204.60 to 205.25 nm, with an optimum pump wavelength of 204.97 nm. The full width at half maximum (FWHM) of the curve is approximately 0.65 nm (154 cm$^{-1}$), about 30 times broader than the 0.02 nm (~5 cm$^{-1}$) FWHM previously observed with nanosecond laser pumping [2]. This is mostly due to the broad bandwidth of the 125 fs pump laser pulse, allowing different frequencies with the same spectral phase to collectively contribute to the two-photon excitation.

Measuring the pump energy dependence of the lasing intensity provides an excellent mean of establishing the presence of lasing. To control the pulse energy of the pump laser, a variable attenuator was inserted inside the frequency-mixing apparatus. The intensity of both the forward and backward 656 nm lasing energies versus pump laser pulse energy is plotted in Fig. 3, in which each point corresponds to the average of 10 laser shots. A power function $E_{lasing} = a + b \times (E_{pump})^c$ is fitted to the experimental data (dashed curve in the figure) where $a, b$ are coefficients taking into account the presence of a lasing threshold, and $c$ is the power index. It is found that the forward lasing energy scales as $2.1 \pm 0.5$ power of the pump pulse energy while the backward lasing energy scales as $2.0 \pm 0.2$. A threshold energy of approximately 5.5 μJ was determined for both lasing directions. This result is firmly consistent with the expected $E_{pump}^2$ dependence of a two-photon process. For femtosecond two-photon excitation, it has been found that interferences due to photodissociation of other species, such as water, CH$_3$, OH etc., are virtually eliminated [15]. Therefore, only the H atoms that are naturally present in the methane/air flame contribute to the lasing signal, which would facilitate quantification.

Aiming to apply the lasing effect for single-ended combustion diagnostics, we are particularly interested in the backward lasing signal. Figure 3 shows a single-shot far-field image recorded with an intensified CCD camera 1.24 m away from the excitation region in the backward direction. As can be seen, the 656 nm lasing beam has a strong donut-shaped spatial mode, surrounded by a much weaker diffracted mode that was not fully detected due to the limited size of the prism. Considering the excitation region as a pencil-shaped cylindrical emitter, its Fresnel number can be calculated as $F = \pi \omega^2 / L\lambda \approx 24$, where $\omega = 100 \ \mu m$ is the transverse radius, $L = 2 \ mm$ is the longitudinal length of the excitation volume and $\lambda = 656 \ nm$ is the lasing wavelength. Since $F \gg 1$, several diffraction-limited modes can be sustained with the geometrical angle $\theta = w/L = 0.05$ [16], as shown in Fig. 4(a). By neglecting the weak surrounding modes, the divergence of the backward 656 nm lasing beam varies with the pump laser energy such that higher pulse energy results in larger divergence.

With a pump pulse energy of 20 µJ, the divergence was determined to ~17 mrad, which is fairly consistent with diffraction-limited lasing from the cross-sectional size of the excitation volume.

With a streak camera, the temporal profile of the backward 656 nm lasing pulse is measured as shown in Fig. 4(b). The result of the single-shot measurement was fitted with a Gaussian function, suggesting a pulse duration of 11 ps. It is noted that the pulse duration fluctuates around an averaged value 15 ps because of slight variations in the pump pulse energy. While experiments with nanosecond laser pumping result in lasing pulses of nanosecond duration exhibiting complicated and spiky temporal structures [17–19], our experiments based on femtosecond laser pumping result in lasing pulses of 15 ps duration with smooth temporal profiles, which are very attractive features for spatially resolved single-ended concentration measurements.

Considering two spatially-separated excitation volumes of hydrogen atoms, these two excitation volumes should give rise to two backward lasing pulses, temporally separated if the lasing pulse durations are sufficiently shorter than the time it takes for a lasing pulse to travel the distance between the two excitation volume. To test this idea, two welding torches (nozzle diameter ~1.5 mm) were used to establish two $CH_4/O_2$ flames ($\phi = 1.0$) with a separation of 13.5 mm. As shown in Fig. 5, the temporal profile of the backward 656 nm lasing, measured by the streak camera, clearly shows two temporally separated pulses, reflecting the presence of two spatially separated flames. The temporal separation is approximately 55 ps, which corresponds to a spatial separation of 16.5 mm. The 3 mm over-prediction is most likely caused by accumulated group velocity delay of both the 205 nm femtosecond pulse and backward 656 nm picosecond lasing pulse trough the two flames. This result indicates that the backward lasing technique offers measurements of hydrogen with a spatial resolution in the millimeter range. The highest spatial resolution obtained in these initial tests is ~7 mm, limited by the minimum separation of two welding torches. A better spatial resolution of the backward lasing technique can be expected. Since resonant two-photon LIF has been observed in O, N, C, CO and $NH_3$, which all are important species in combustion processes, femtosecond two-photon excited backward lasing of these species can in principle be observed under certain experimental conditions. Thus, it appears feasible to extend the present measurement concept towards single-ended spatially resolved detection of these species as well.

**Summary**

In summary, we report on an observation of backward lasing effect of atomic hydrogen in a premixed methane/air flame using femtosecond deep-UV laser pulses. Following femtosecond resonant two-photon excitation, 656 nm lasing emission occurs in both the forward and backward direction. Characterization of the backward lasing pulse was carried out, including analysis of the emission spectrum, pump energy dependence, spatial and temporal profiles. The duration of the backward lasing pulse was found to be approximately 15 ps, suggesting a strong potential for spatially resolved measurements of atomic hydrogen in the millimeter rang in flames. Using two separated methane/oxygen flames burning on two welding torches, we successfully managed to obtain two backward 656 nm lasing pulses on a streak camera, with a temporal separation consistent with the spatial separation between the two flames, and a best spatial resolution of approximately 7 mm. Based on these results, we believe that the backward lasing effect holds great potential for single-ended concentration measurements, which would constitute a very powerful tool for combustion diagnostics in intractable geometries with limited optical access.

**Funding.** This work was funded through grants from the Knut and Alice Wallenberg Foundation, the Swedish Energy Agency via the Center for Combustion Science and Technology (CECOST), and the ERC (an advanced grant, project: TUCLA).

**Acknowledgement.** The authors would like to thank Dr. Andreas Ehn for fruitful discussions of spatially resolved measurements.

**Reference**

1. M. Aldén, U. Westblom, and J. E. Goldsmith, "Two-photon-excited stimulated emission from atomic oxygen in flames and cold gases.," Opt. Lett. **14**, 305–307 (1989).
2. J. E. M. Goldsmith, "Two-photon-excited stimulated emission from atomic hydrogen in flames," J. Opt. Soc. Am. B **6**, 1979–1985 (1989).
3. M. Aldén, P.-E. Bengtsson, and U. Westblom, "Detection of carbon atoms in flames using stimulated emission induced by two-photons laser excitation," Opt. Commun. **71**, 263–268 (1989).
4. S. Agrup, U. Westblom, and M. Aldén, "Detection of atomic nitrogen using two-photon laser-induced stimulated emission: Application to flames," Chem. Phys. Lett. **170**, 406–410 (1990).
5. U. Westblom, S. Agrup, M. Aldén, H. M. Hertz, and J. E. M. Goldsmith, "Properties of laser-induced stimulated emission for diagnostic purposes," Appl. Phys. B **50**, 487–497 (1990).

6. M. A. Nikola Georgiev, Kaj Nyholm, Rolf Fritzon, "Developments of the amplified stimulated emission technique for spatially resolved species detection in flames," Opt. Commun. **108**, 71–76 (1994).
7. Q. Luo, W. Liu, and S. L. Chin, "Lasing action in air induced by ultra-fast laser filamentation," Appl. Phys. B Lasers Opt. **76**, 337–340 (2003).
8. A. Dogariu, J. B. Michael, M. O. Scully, and R. B. Miles, "High-gain backward lasing in air.," Science **331**, 442–445 (2011).
9. P. R. Hemmer, R. B. Miles, P. Polynkin, T. Siebert, A. V Sokolov, P. Sprangle, and M. O. Scully, "Standoff spectroscopy via remote generation of a backward-propagating laser beam.," Proc. Natl. Acad. Sci. U. S. A. **108**, 3130–4 (2011).
10. S. Mitryukovskiy, Y. Liu, P. Ding, A. Houard, and A. Mysyrowicz, "Backward stimulated radiation from filaments in nitrogen gas and air pumped by circularly polarized 800 nm femtosecond laser pulses," Opt. Express **22**, 12750 (2014).
11. P. Ding, S. Mitryukovskiy, A. Houard, E. Oliva, A. Couairon, A. Mysyrowicz, and Y. Liu, "Backward Lasing of Air plasma pumped by Circularly polarized femtosecond pulses for the saKe of remote sensing (BLACK)," Opt. Express **22**, 29964 (2014).
12. Y. Liu, P. Ding, G. Lambert, A. Houard, V. Tikhonchuk, and A. Mysyrowicz, "Recollision-Induced Superradiance of Ionized Nitrogen Molecules," Phys. Rev. Lett. **115**, 133203 (2015).
13. P. Ding, E. Oliva, A. Houard, A. Mysyrowicz, and Y. Liu, "Lasing dynamics of neutral nitrogen molecules in femtosecond filaments," Phys. Rev. A - At. Mol. Opt. Phys. **94**, 1–6 (2016).
14. P. Ding, J. C. Escudero, A. Houard, A. Sanchis, J. Vera, S. Vicéns, Y. Liu, and E. Oliva, "Nonadiabaticity of cavity-free neutral nitrogen lasing," Phys. Rev. A **96**, 33810 (2017).
15. W. Kulatilaka, J. Gord, V. Katta, and S. Roy, "Photolytic-interference-free, femtosecond two-photon fluorescence imaging of atomic hydrogen," Opt. Lett. **37**, 3051–3053 (2012).
16. M. Gross and S. Haroche, "Superradiance: An essay on the theory of collective spontaneous emission," Phys. Rep. **93**, 301–396 (1982).
17. S. K. R.C.Y. Auyeung, D.G. Cooper, and B. Feldman "Stimulated emission in atomic hydrogen at 656 nm," Opt. Commun. **79**, 7–10 (1990).
18. S. Agrup and M. Aldén, "Two-photon laser-induced fluorescence and stimulated emission measurements from oxygen atoms in a hydrogen / oxygen flame with picosecond resolution," Opt. Commun. **113**, 315–323 (1994).

19. A. J. Traverso, R. Sanchez-Gonzalez, L. Yuan, K. Wang, D. V Voronine, A. M. Zheltikov, Y. Rostovtsev, V. Sautenkov, A. V Sokolov, S. W. North, and M. O. Scully, "Coherence brightened laser source for atmospheric remote sensing.," Proc. Natl. Acad. Sci. U. S. A. **109**, 15185–90 (2012).

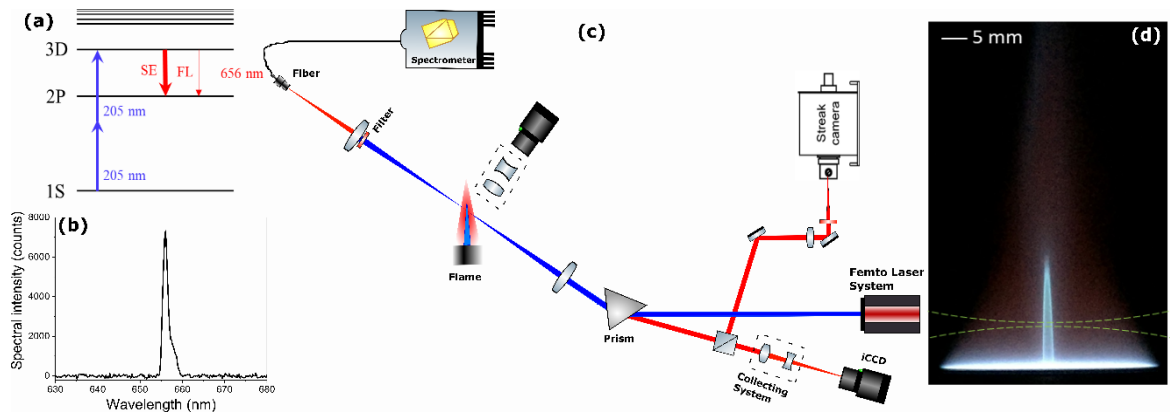

Fig. 1. (a) Energy levels of atomic hydrogen relevant to 205 nm-two-photon-excited fluorescence (FL) and stimulated emission (SE). (b) Spectrum of backward 656 nm lasing pulse of atomic hydrogen in premixed methane/air flame, recorded with a fiber spectrometer (Avantes). (c) Schematic illustration of the experimental setup. (d) A flame-brightness image of the premixed methane/air flame, which clearly shows a much brighter jet flame in the center surrounded by a less bright outer flat flame. The dashed line represents focusing of the 205 nm femtosecond laser pulses.

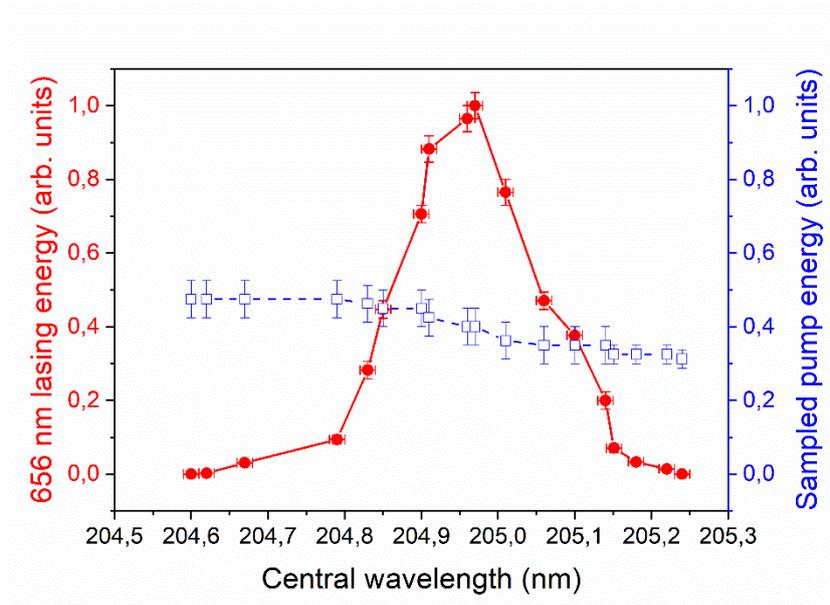

Fig. 2. Pump wavelength scan of the femtosecond two-photon-excited 656 nm lasing of atomic hydrogen by using ~20 µJ-energy pulses in a rich (ϕ =1.3) methane/air flame. The open square curve shows the simultaneously sampled pump laser energy.

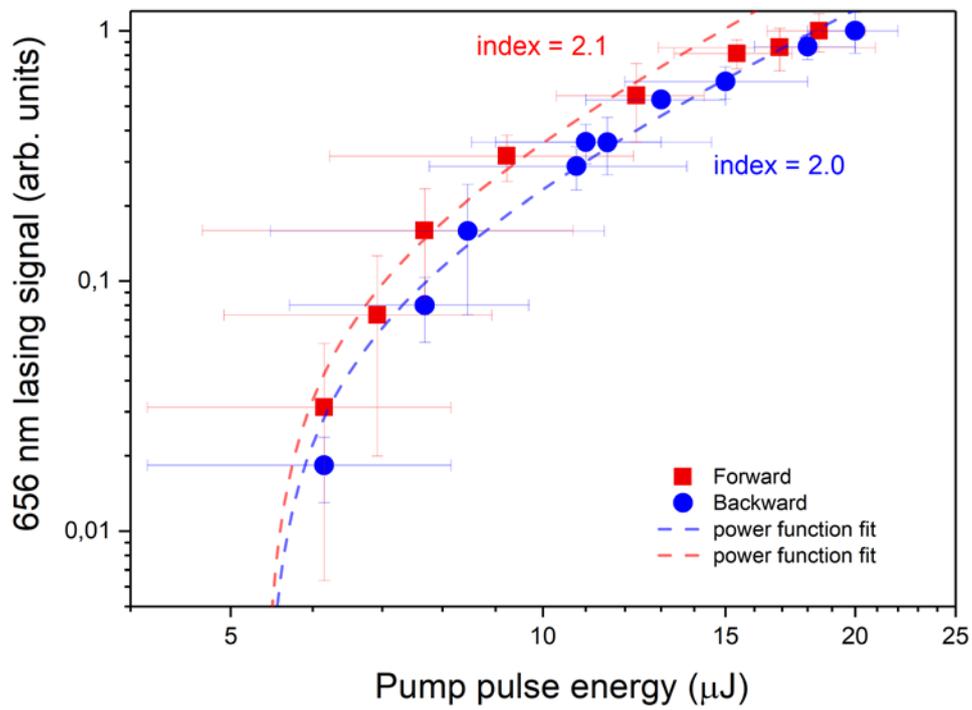

Fig. 3. Pump energy dependence of both the forward and backward 656 nm lasing signals. The dashed curves represent the power function fittings with power indices of 2.1 and 2.0, respectively.

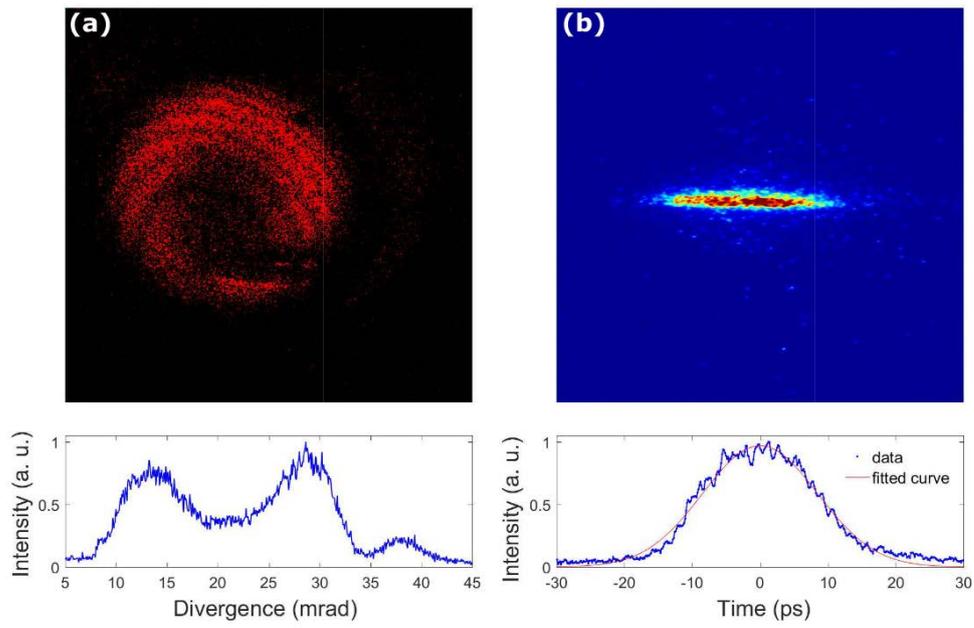

Fig. 4. Single-shot spatial (a) and temporal (b) profile of the backward 656 nm lasing pulse of atomic hydrogen generated in a premixed ϕ=1.3 methane/air flame.

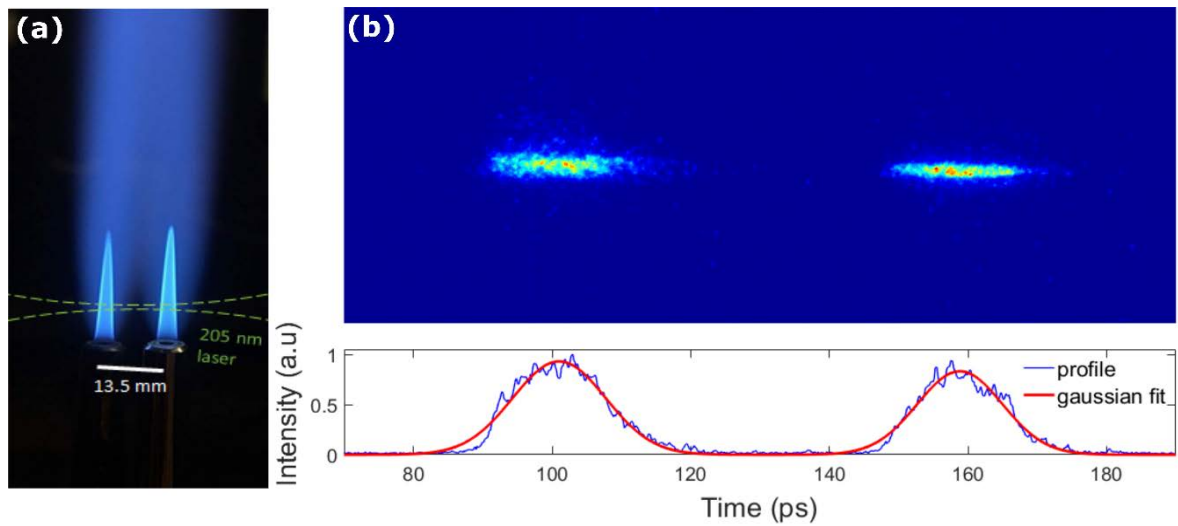

Fig. 5. (a) Image of flames burning on two welding torches, with illustration of the focusing of the 205 nm laser beam in the center between the two flames. (b) Temporal profile of the backward 656 nm lasing signal recorded by a streak camera. The red curve represents a Gaussian fit, suggesting pulse durations of 9.5 and 8.7 ps for the two lasing pulses respectively.